\documentclass[aps,prl,preprint,groupedaddress,showpacs,showkeys]{revtex4}
\usepackage{bm}

\begin{document}

%\preprint{}

\title{Speed of gravity and gravitomagnetism}


\author{J.-F. Pascual-S\'anchez}
\email[Electronic address: ]{jfpascua@maf.uva.es}
%\thanks{}
\affiliation{Dept. Matem\'atica Aplicada, Facultad de Ciencias,
Universidad de Valladolid, 47005, Valladolid, Spain, EU}


%\date{\today}

\begin{abstract}
A $v_J/c$ correction to the Shapiro time delay  seems verified by
a 2002 Jovian  observation by VLBI. In this Essay, this correction
is interpreted  as an effect of the aberration of light in an
optically refractive medium which supplies an analog of Jupiter's
gravity field rather than as a measurement of the speed of
gravity, as it was first proposed by other authors. The variation
of the index of refraction is induced by the Lorentz invariance of
the weak gravitational field equations for Jupiter in a uniform
translational slow motion with velocity $v_J=13.5\, km/s$. The
correction on time delay and deflection is  due not to the Kerr
(or Lense-Thirring) stationary gravitomagnetic field of Jupiter,
but to its Schwarzschild gravitostatic field when measured from
the barycenter of the solar system.
\end{abstract}


\maketitle



\section{Introduction.}
In general relativity, the causal propagation of gravity in
vacuum, in the sense that it cannot travel faster than light in
vacuum, usually follows from the hyperbolic form of the Einstein
field equations in empty spacetime. Moreover, a general purely
geometrical proof of this fact is given in  \cite{sen} and it does
not seem to be a direct consequence of the hyperbolic character of
Einstein's theory.

In spite of that, three years ago Kopeikin \cite{1Kop} suggested a
solar system time delay test  from which he concluded that the
speed of gravity $c_g$ could be directly measured and hence
compared with the speed of light $c$ in vacuum. His final  result
was $c_g= (1.06 \pm 0.21)\,c$. A more detailed description of this
solar system time delay test is the following: In Ref.\cite{1Kop}
an extra time delay of the quasar light was predicted, caused by
the passage of Jupiter by the quasar QSO J0842+1835 at a
separation of only 3.7 arcminutes (14 Jovian radii) on September
8, 2002, which was measured by advanced VLBI and recorded by
eleven radio antennae \cite{Fom}, an impressive observational
feat.  The observation was conducted by the National Radio
Astronomical Observatory (USA) and the Max Planck Institute for
Radio Astronomy (EU).

The extra time delay is the total time delay, due to the
difference in the arrival time of two light rays leaving the
quasar simultaneously, minus the classical Shapiro time delay
\cite{Shap} caused by the Schwarzschild gravitational field of a
static  mass at rest. As Jupiter's gravity is very weak, the
Jovian time delay (or deflection) observation measurement required
remarkable accuracy: few picoseconds (or about 10 $\mu$arcsec).
The order of magnitude of the time delay prediction is a delay of
115 psec (deflection of 1190 $\mu$arcsec) for the radial (static)
term, and a time delay of 4.8 psec (deflection of 51 $\mu$arcsec)
for the transverse (velocity) term at the point of the closest
approach. Both these predicted theoretical values have been
successively reduced from Ref.\cite{1Kop} to \cite{Fom, Kop1}.

In a series of papers the authors of
\cite{1Kop,Fom,Kop1,kopeikin,kop1,kop2} have argued that the
additional terms in  the time delay depend upon the propagation
speed of gravity, giving the aforementioned final result
$c_g=(1.06 \pm 0.21)\,c$. Actually, the authors of
\cite{1Kop,Fom,Kop1,kopeikin,kop1,kop2} use a kind of
vector-tensor theory instead of the Einstein's general relativity
which is a pure tensor theory. So far, different criticisms from
various authors \cite{will,asa} have inhibited the acceptance of
this interpretation of the experiment although, seemingly, they
have not ruled it out.

The main purpose of this Essay is to show an altogether new
approach which, in our opinion, completely invalidates  the claim
made in \cite{kopeikin,kop2}. In these works it was argued that,
working in the linear weak field approximation, a gravitomagnetic
field appears after a passive Lorentz transformation from the
static rest frame of Jupiter to the barycenter of the solar system
has been applied. In \cite{kopeikin,kop2}, the gravitomagnetic
field of Jupiter is the cause of the extra delay of the
quasarlight by dragging it in the direction of motion of Jupiter.

Here, we will show that a intrinsic gravitomagnetic field does not
appear after making the Lorentz transformation. Hence, the extra
time delay effect is only a test of the Lorentz invariance of the
weak gravity equations and it is actually a fine measurement with
VLBI of the aberration of light already observed by Bradley in
1728 using a telescope.

\section{Criterion for an intrinsic gravitomagnetic field. }
At first sight, the linearized theory of general relativity or
gravitomagnetism may be thought of as that phenomenon by which the
spacetime geometry and its curvature change. This change is due to
 mass-energy density and mass-energy currents relative to other
mass-energy.

It is possible to use the analogy with electromagnetism as in
\cite{Mash}, however, we stress that apart from formal analogies,
gravitomagnetism and the Maxwell-Lorentz electromagnetic theory
are fundamentally different. The main difference is the
equivalence principle: in freely falling frames  it is possible to
eliminate the local acceleration effects (first derivatives of the
metric $ g_{\alpha\beta}$) of a gravitational field.  To
characterize the electromagnetic field one must calculate the two
Casimir invariants under the Poincar\'e group of the Faraday
tensor $F_ {\alpha \beta}$. They are the scalar, $-{1 \over 2} F_
{\alpha \beta} F^ {\alpha \beta} = \|\bm{E}\|^2 - \|\bm{B}\|^2 $,
and the pseudoscalar, ${1 \over 4} F_{\alpha \beta} \,^\star
F^{\alpha \beta} = \bm{E} \cdot \bm{B}$, where ``$^\star$" is the
dual operation. If we have only a charge $q$, in the rest frame
``$\bm{r}$", then we have an electric field $\bm{^r\! E} \not = 0$
but the magnetic one is $\bm{^r\! B}=0$ and the invariant $F_
{\alpha \beta} \,^\star F ^{\alpha \beta}$ is zero. Therefore,
even in other inertial frames where $\bm{B} =- \gamma \, (\bm{v}/c
\times \bm{^r\! E}) \not = \bm{0}$ and $\bm{E} \not = \bm{0}$,
this pseudoinvariant will be zero. However, if in the rest frame
we have a charge $q$ and a magnetic dipole $\bm{m}$ then
$F_{\alpha \beta } \,^ \star F^{\alpha \beta } \not = 0 $ and this
pseudoinvariant will be different from zero in any other inertial
frame.

Something similar occurs in general relativity. Now, only the
diagonal components of the Schwarzschild metric are non-vanishing
when these are measured in the rest frame ``$\bm{r}$''. These
components are the ``gravitoelectric'' $g_{00}$ and ``space's
curvature" $g_{ij}$ potentials. However, if we consider a boosted
inertial frame with velocity, $\bm{v}=v_i\,e^i$, $i=1,2,3$,
relative to the mass $M$, we will have the ``gravitomagnetic''
components $g_{0 i}= -4 GM v_i/ rc^3$, proportional to the
velocity of the source measured by an observer of the moving
inertial frame. In this case we have an extrinsic gravitomagnetic
field.

The metric for this observer  reads at linear approximation as:
\begin{equation}{\label{1}}
ds^2 = - (1 + 2 U/c^2  ) c^2\, dt^2 + (1  - 2 U/c^2 )\, d\bm{x}
\,d\bm{x} \, + 2 g_{0i} c\,dt\, d\bm{x},
\end{equation}
where $U=-GM/r$ is the Newtonian potential. In every spacetime we
can always make the ``gravitomagnetic" vector potential $\bm{A}$,
defined by $ g_ {0i}=-2 A_i/c^2$, different from zero, realizing a
coordinate transformation. Hence, to study the intrinsic
properties of a gravitational field one has to compute the
invariants of the Riemann curvature tensor instead of analizing
the metric components in a particular coordinate system.

In gravitational theory, the invariants of the curvature tensor
analogous to the electromagnetic ones are the Kretschmann
invariant $ R_ {\alpha \beta \mu \nu } R^ {\alpha \beta \mu \nu }$
and the pseudo-invariant $ ^\star R ^{\alpha \beta\mu \nu } R _
{\alpha \beta \mu \nu}= \,^\star \bm{R} \cdot \bm{R}$. Hence, if
one only considers a  mass  $M$ (the Schwarzschild solution) then,
in the rest frame ``$\bm{r}$" of this mass, one has that the
``gravitoelectric" components $ \bm{ ^r}R_{i0j0}$ are non-zero
whereas $^\star \bm{R} \cdot \bm{R}$ is zero.

Due to the Lorentz invariance of the weak gravitational equations,
this also happens if the invariants are measured from other
inertial frames where  the ``gravitomagnetic"  components are
$R_{i0jk} \not= 0$. Then at the lowest order, the pseudo-invariant
is
 $^\star \bm{R} \cdot \bm{R} \, \simeq
\nabla^2 \, [\nabla (U/c^2)\cdot (\nabla \wedge \,\bm{A}/c^2 )] $.
For a static Schwarzschild metric boosted with velocity $\bm{v}$,
we have $\bm{A}= -2 U\bm{v}/c$. However, $^\star \bm{R} \cdot
\bm{R}$ is still zero:
\begin{equation}{\label{2}}
 ^\star \bm{R} \cdot \bm{R} \, \simeq \nabla^2 \,
[\nabla (U/c^2) \cdot (\nabla\wedge U\bm{v}/c^3 )] = 0.
\end{equation}
In this case, the ``gravitomagnetic" components $R_{i0jk}$ just
measure the effect of a (passive) local Lorentz transformation
from the rest frame of the mass $M$ to a boosted frame on a static
gravitational field which remains intrinsically unchanged. Hence,
an intrinsic gravitomagnetic field does not appear in the boosted
linear Schwarzschild metric (\ref{1}).

However, if one considers a steady rotating source with mass $M$
and intrinsic angular momentum $\bm{S}$ (that is to say, Kerr
solution, or its linear approximation: Lense-Thirring), the
gravitomagnetic potential is $\bm{A} = G \,\bm{ S} \wedge \bm{r}/
c\,r^3$, and \textit{ both curvature invariants are non-zero in
any inertial frame}. The last sentence is the criterion of the
existence of a intrinsic gravitomagnetic field. Note that if this
criterion is not satisfied then some confusion might appear
concerning the gravitomagnetic properties of a gravitational
field. For instance,  a geodetic de Sitter effect might be
misinterpreted as an intrinsic gravitomagnetic Lense-Thirring one
due to the orbital angular momentum $\bm{L}$  measured by a
orbiting observer freely falling around a non-rotating mass.

\section{Interpretation of the 2002 Jovian observation.} According
to Einstein's theory, a gravitational field is identified with the
spacetime curvature. Also, in an analogous way, it can be
considered as a retarder, a deflector and a lens simultaneously,
i.e., as an inhomogeneous refractive medium of index $n(r)$ acting
upon the propagation of the light rays along a trajectory between
two points in spacetime. Moreover, an optical refractive medium
causes a time delay (proportional to $n$), a deflection of the
light rays (through the spatial gradient of $n$) and the
appearance of lensed images. In general, these three actions are
mathematically related through the phase or eikonal function and
its successive derivatives. However, the first two effects can
also take place in the Minkowski spacetime, when gravitation is
neglected, due to the Doppler or the aberration special
relativistic kinematical effects. In this last case both effects
are measured at the same point by two different observers.

The effect of gravity is to make the medium optically denser in
the vicinity of a mass and hence the coordinate speed of light
diminishes as we approach the mass and  we could say that the
light is repelled by the mass. However, let us remark that the
invariant speed of light measured by any observer will be always
$c$, as in vacuum. For a frame at rest with a static mass,
described by the linear Schwarzschild metric,
 the index is $\bm{ ^r}n(r)=1-2U/c^2$. When it is  measured from a boosted
inertial frame, the index is
\begin{equation}{\label{3}}
n(r)=\left[\bm{ ^r}n(r)-\frac{v}{c}\,\right]\left[{1-\frac{\bm{
^r}n(r)\,v}{c}}\right]^{-1},
\end{equation}
which at first order w.r.t. $v/c$ and $U/c^2$, gives the
refractive index associated with the boosted linear Schwarzschild
metric (\ref{1}), (as in Ref. \cite{Fri}), which reads
\begin{equation}{\label{4}}
n(r)=1-2U/c^2-4(v/c)U/c^2,
\end{equation}
where $n(r)>\,\bm{ ^r}n(r)$.

On the other hand, following \cite{1Kop,Kop1,will,Will93}, the
total (Shapiro plus extra) time delay caused by  Jupiter's gravity
and measured by two VLBI  antennae can be written as
\begin{equation}{\label{5}}
\Delta T= \frac{2GM_J}{c^3}\,(1 - \frac{\bm{k} \cdot \bm{v_J}}{c})
\, \ln \frac{r_{1J} - \bm{K} \cdot  \bm{r_{1J}}}{r_{2J} - \bm{K}
\cdot  \bm{r_{2J}}},
\end{equation}
where  $\bm{k}$ is the unit vector from the quasar to the
barycenter of the solar system; $M_J$ and $\bm{v_J}$ are the
Jovian mass and velocity; $\bm{r_{iJ}}$ are the difference between
the barycenter coordinates of the {\it i}-th VLBI station and
Jupiter and, finally, $\bm{K}$ is given by $\bm{k} -
\frac{1}{c}\bm{v_{TJ}}$, where $\bm{v}_{TJ}=\bm{k} \wedge
(\bm{v_J}\wedge \bm{k})$ is the transverse Jovian velocity in the
plane of the sky which supplies the aberration of light. In
(\ref{5}) two different corrections $\frac {1}{c}\bm{v_J}$ appear,
the first one in the  logarithmic term and the second  in the
pre-factor. In the considered observation  the main one is the
correction, from $\bm{k}$ to $\bm{K}$ due to
$-\frac{1}{c}\bm{v_{TJ}}$, in the argument of the logarithmic
term. All quantities are evaluated at the same time $t$ of
reception of the light on Earth.

Since the above formula (\ref{5}) is obtained by means of a
Lorentz transformation from the rest frame of Jupiter (where
Shapiro's formula is applied) to the barycenter frame of the solar
system then, by the arguments exposed in this letter, a intrinsic
gravitomagnetic field does not exist. In spite of that, in
\cite{1Kop,Fom,Kop1,kopeikin,kop1,kop2} the gravity speed $c_g$,
due to Jupiter's gravitomagnetic field, appears in the above
formulas instead of the speed of light $c$. We conclude that {\it
the right interpretation of the $v_J/c$ correction terms in the
time delay involves, as a result of a passive Lorentz
transformation, the aberration of the speed of light rather than
the speed of gravity}.

There are two main statements  about the gravitomagnetic field of
Jupiter in \cite{kopeikin,kop2}. The first one appears in
\cite{kopeikin} and says: ``Another general relativistic
interpretation  of the Jupiter-quasar experiment (apart from its
association with the measurement of the speed of gravity) consists
in the statement that the experiment has measured the magnitude of
the gravitomagnetic field generated by the orbital motion of
Jupiter''. The second statement appears  in \cite{kop2} and
reads: "Thus, in a moving frame, the translational motion of
Jupiter produces the gravito-magnetic field which deflects the
light by dragging it to the direction of motion of Jupiter. We can
measure this gravito-magnetic dragging of light and express its
magnitude in terms of the speed of characteristics of...".
However, we think that our above analysis shows that the
interpretation given in the quoted statements cannot be applied to
the 2002 Jovian observation.

\section{acknowledgments}
I wish to thank C. Dehesa and J. Mart\'{\i}n  for discussions on
this subject and to A. San Miguel, M. del Mar San Miguel and F.
Vicente for comments and a critical reading of the manuscript. A
previous version of this work received an Honorable Mention in the
2004 Essay Competition of the Gravity Research Foundation (GRF)
and it has appeared in the arXiv as gr-qc/0405123, before
Ref.\cite{kopeikin} was revised and published as \cite{kop1}. The
author is currently partially supported by MCYT TIC2003-07020.
%\end{document}


\end{document}